\documentclass[
twocolumn,
]{ceurart}

\usepackage{graphicx}
\usepackage{xspace}
\usepackage{subcaption}

\newcommand*{\GARLIC}{\texttt{GARLIC}\xspace}

\newcommand*{\ONION}{\texttt{ONION}\xspace}
\usepackage{listings}
\begin{document}

\copyrightyear{2025}
\copyrightclause{Copyright © 2025 for this paper by its authors. Use permitted under Creative Commons License Attribution 4.0 International (CC BY 4.0).}
\conference{DataEd'26: 5th International Workshop on Data Systems Education}

\title{Seasoning Data Modeling Education with GARLIC: A Participatory Co-Design Framework}

\tnotemark[1]
\tnotetext[1]{Published in the Proceedings of the Workshops of the EDBT/ICDT 2026 Joint Conference (March 24-27, 2026), Tampere, Finland}


\author[1]{Viktoriia Makovska}

\author[2]{Ihor Michurin}

\author[2]{Mariia Tokhtamysh}

\author[3]{George Fletcher}

\author[4]{Julia Stoyanovich}

\address[1]{Ukrainian Catholic University, Ukraine}
\address[2]{Kharkiv National University of Radioelectronics, Ukraine}
\address[3]{Eindhoven University of Technology, The Netherlands}
\address[4]{New York University, USA}

\begin{abstract}
Entity–Relationship (ER) modeling is commonly taught as a primarily technical activity, despite its central role in shaping how data systems represent people, processes, and institutions. Prior research in participatory design demonstrates that involving diverse stakeholders in modeling can surface tacit knowledge, challenge implicit assumptions, and produce more inclusive data representations. However, database education currently lacks structured pedagogical approaches for teaching participatory ER modeling in practice.

We introduce the \GARLIC 
methodology for teaching and learning participatory ER modeling. \GARLIC adapts and extends the \ONION participatory ER modeling framework of Makovska et al.\ (HILDA 2025) into a workshop-based learning format that combines role-playing, collaborative synthesis, guided critique, and iterative refinement. \GARLIC is designed to develop both technical modeling skills and critical awareness of the social and ethical dimensions of data representation. \GARLIC lowers the barrier to participatory ER modeling and equips students with practical skills for collaborative, inclusive data model design.
\end{abstract}


\maketitle

\section{Introduction}


Participatory design (PD) \cite{PD_paper} has become an important orientation in human–computer interaction, responsible AI, and sociotechnical systems research. A central insight across these areas is that design decisions are never neutral: they embed values, assumptions, and power relations that shape how systems operate and whom they serve. Design justice research shows that traditional, expert-driven design practices often privilege dominant perspectives while marginalizing the communities most affected by data-intensive systems \cite{costanza2020design}. PD emerged in response to these concerns, emphasizing shared decision-making and the inclusion of lived experience in design processes \cite{schuler1993pd}. Recent work in participatory and responsible AI further demonstrates that data modeling choices can have ethical, political, and societal consequences that extend far beyond technical considerations \cite{birhane2022participatoryAI, fletcherHILDA2025,suresh2024participationFoundationModels}.

Despite these insights, Entity–Relationship (ER) modeling is still predominantly taught as a technical activity focused on correctness and syntactic validity. Students are typically trained to produce schemas that are structurally sound, with less attention to stakeholder interpretation, contextual fit, or the implications of representational choices. 
Indeed, popular data management textbooks, such as \citet{Silberschatz2020} and \citet{Elmasri2016}, focus heavily on technical correctness. They primarily teach students how to fix structural errors through normalization and how to improve performance. However, these resources rarely teach the social side of modeling. They present `requirements gathering' as a simple first step where the architect collects facts, rather than a complex dialogue with stakeholders.

Computer science curricula further reinforce this separation. ACM/IEEE recommendations \cite{ACM2013, ACM2023} typically address social issues in standalone ethics modules. While these help students critique unfair systems, they do not teach how to design inclusive data models. Students learn ER diagram syntax but not how to address 
the mismatch between real-world complexity and database schemas \citep{onion}.

Recently the \ONION framework for partcipatory ER design was introduced \cite{onion}.
\ONION demonstrates that involving stakeholders can surface tacit knowledge and reveal misalignments that experts alone may overlook, but it does not address how future engineers should be taught to facilitate such participatory processes. Indeed, \ONION
was designed for professional design contexts rather than for education.

As a result, a clear educational gap remains. There is no structured pedagogy that teaches participatory ER modeling as a practical skill: eliciting diverse perspectives, integrating them into the modeling process, and validating their representation in the final schema. Consequently, database education remains largely disconnected from participatory and sociotechnical design practices.

To address this gap, we introduce \GARLIC (Generalized, Accessible, RelationaL, Inclusive Co-Design), an educational methodology for teaching participatory ER modeling through a structured, workshop-based approach. \GARLIC brings participatory principles into the classroom using scenario framing, role-based perspective taking, collaborative synthesis, and explicit validation of participant perspectives in the resulting ER model. 
\GARLIC provides students with a concrete process for experiencing how perspectives shape data models and how participatory success can be evaluated. 

In this paper we present the results of our work in progress on developing \GARLIC.  All current materials and instructions for reproducing the \GARLIC workshop are publicly available.\footnote{\url{https://github.com/mariykadreams/GARLIC}}

\section{Prior Work}

Designing database systems involves both social and technical factors. Traditional methods have focused primarily on technical factors such as scalability and efficiency. There is a growing awareness, motivated by Design Justice~\cite{costanza2020design} and participatory frameworks, that social context and inclusivity should be central to the design process. However, existing research has not applied these socio-technical principles to the teaching of database modeling. This section explains how our work relates to prior research,  drawing on studies in participatory design to highlight the importance of our investigation.

\paragraph*{Traditional vs.\ Participatory ER Modeling}

Current ER modeling methodologies are predominantly technocentric, focusing on the formalization of requirements by technical experts. While effective in terms of system performance, these ``expert-only'' models often suffer from \emph{semantic gaps}—disconnections between the database schema and the lived realities of stakeholders.

The \ONION framework~\cite{onion} is a multi-layered participatory ER modeling methodology that structures collaboration through five stages: Observe, Nurture, Integrate, Optimize, and Normalize. These stages support a progressive abstraction from unstructured stakeholder narratives to a formal ER diagram, while preserving traceability of stakeholder input throughout the process. ONION was designed for professional and research-oriented participatory design settings, where facilitators and technical experts already possess experience with collaborative design practices. As such, it deliberately abstracts away from pedagogical concerns such as novice cognitive load, instructional scaffolding, and explicit guidance for teaching facilitation skills.

Building on this participatory foundation, the \ONION methodology aims to address semantic gaps by incorporating storytelling and group sketching into the ER modeling process. However, the literature reveals a critical limitation: although \ONION provides a mechanism for involving stakeholders, there is no established method that teaches STEM students how to lead and facilitate participatory data modeling. Most research in database education concentrates on reducing syntactic errors or improving individual conceptual understanding, rather than preparing students to navigate power imbalances and communication challenges inherent in inclusive design


\paragraph*{Challenges in Database Education}

Research in database education consistently identifies Conceptual Modeling (ER modeling) as one of the most difficult tasks for students \cite{Batra1992, Murray2016}. Unlike programming, which provides immediate feedback through compiler errors, data modeling requires abstract thinking that is challenging to validate. Learners often struggle with `cognitive load' when translating ambiguous real-world requirements into rigid database schemas, leading to frequent errors \cite{Batra1994}.

Although education researchers have developed interventions to help, such as using concept maps or `learning-from-errors' approaches \cite{Shet2013}, these solutions usually focus on technical accuracy within the classroom. They treat the data model as a solved puzzle. There is a notable absence of structured toolkits that enable students to apply these skills in multi-stakeholder, participatory contexts. As a result, a gap remains: students learn to build technically valid models, but they do not learn how to validate models with diverse stakeholders such as the people represented by the data.

\paragraph*{Overview of Validation Techniques in Pedagogical Research}

To evaluate the effectiveness of new teaching methods, prior studies typically adopt a mixed-methods validation approach combining qualitative and quantitative measures. An analysis of related work (e.g.,~\cite{birhane2022participatoryAI,suresh2024participationFoundationModels}) reveals three commonly used validation strategies:
    (1)  Measuring improvements in students' technical skills through pre- and post-intervention assessments;
    (2)  Having senior database architects or educators review student-created models for accuracy and completeness;
    (3)  Administering surveys to assess participants' perceived inclusion and the extent to which the models reflect their needs.

Building on these approaches, we outline a validation perspective for the technical quality of the ER models and the design process itself. In particular, we assess students' ability to facilitate inclusive discussions and examine stakeholder satisfaction with the participatory design experience.

\paragraph*{Research Gaps and Significance}

A review of the literature reveals two critical gaps:
    First, we have a \textbf{methodological gap:} no existing curriculum or toolkit provides a step-by-step guide enabling STEM students to grow from `technical experts' to participatory facilitators.
    Second, there is \textbf{the practical educational gap:} while theoretical frameworks for participatory ER modeling exist, they are not inherently pedagogical. \ONION structures participatory ER modeling into five stages (Observe, Nurture, Integrate, Optimize, Normalize), but it does not tell an educator \textit{how} to teach these concepts to novices. Students, who already struggle with the cognitive load of basic ER syntax \cite{Batra1992}, cannot effectively apply complex research frameworks without structured educational scaffolding.

Addressing these gaps is vital. Without meaningful power-sharing, ``participation'' remains symbolic \cite{costanza2020design,Arnstein1969}. While participatory design produces more relevant and equitable systems, students require a structured toolkit to move beyond tokenism. 
\GARLIC operationalizes participatory design within STEM pedagogy to solve a persistent challenge in ER modeling. By bridging these methodological and practical gaps, we prepare engineers to create socially accountable systems, impacting both STEM education and equitable design practice.
\section{A Workshop-Based Approach to Participatory Modeling}

\begin{figure*}[t]
    \centering
    \begin{subfigure}[t]{0.44\textwidth}
        \centering
        \includegraphics[width=\linewidth]{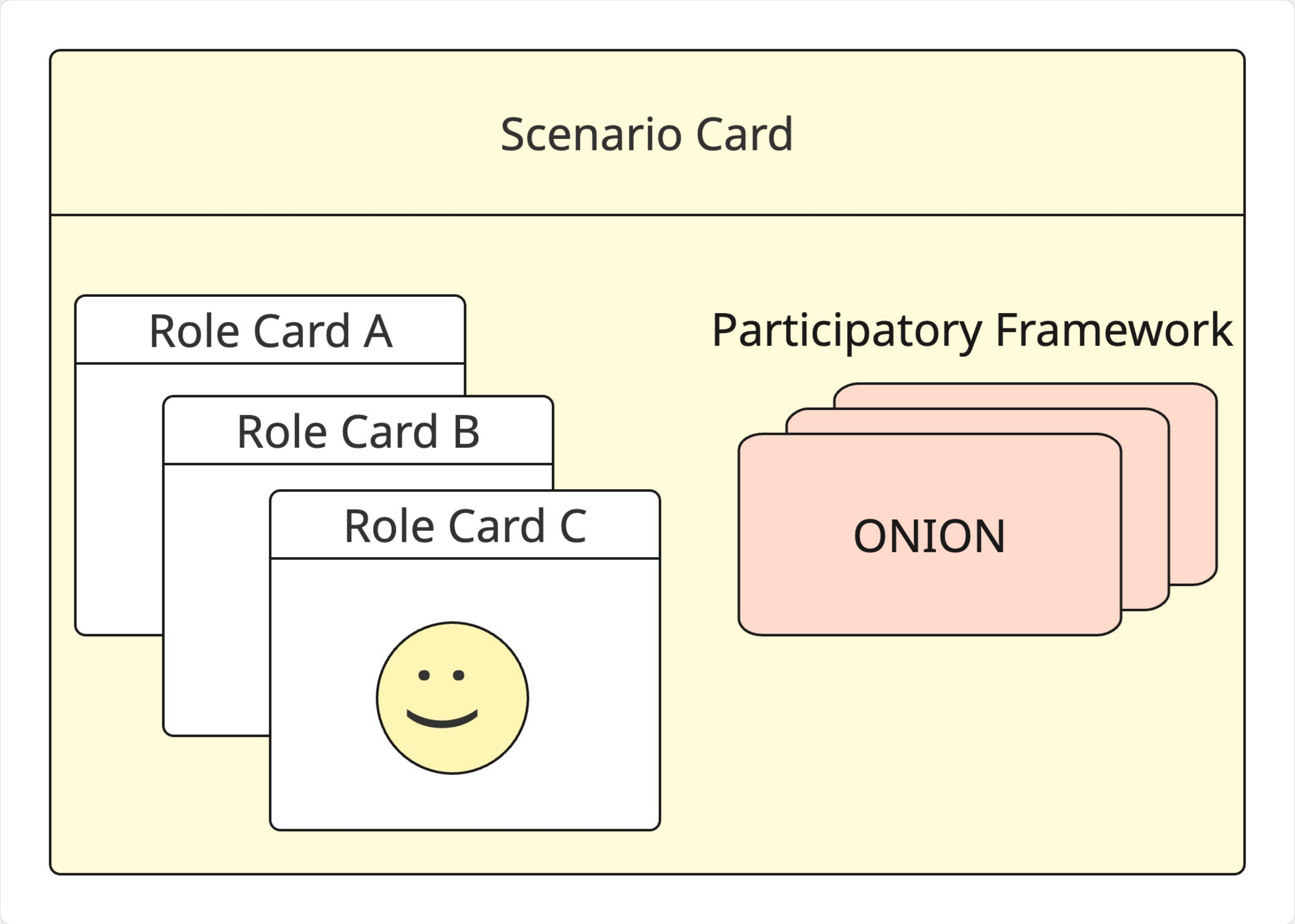}
        \caption{The Scenario Card provides a common context and framing, Role Cards capture 
        individual stakeholder perspectives, and the \ONION framework guides their 
        integration into a shared ER model.}
        \label{fig:method-overview}
    \end{subfigure}
    \hfill
    \begin{subfigure}[t]{0.55\textwidth}
        \centering
        \includegraphics[width=\linewidth]{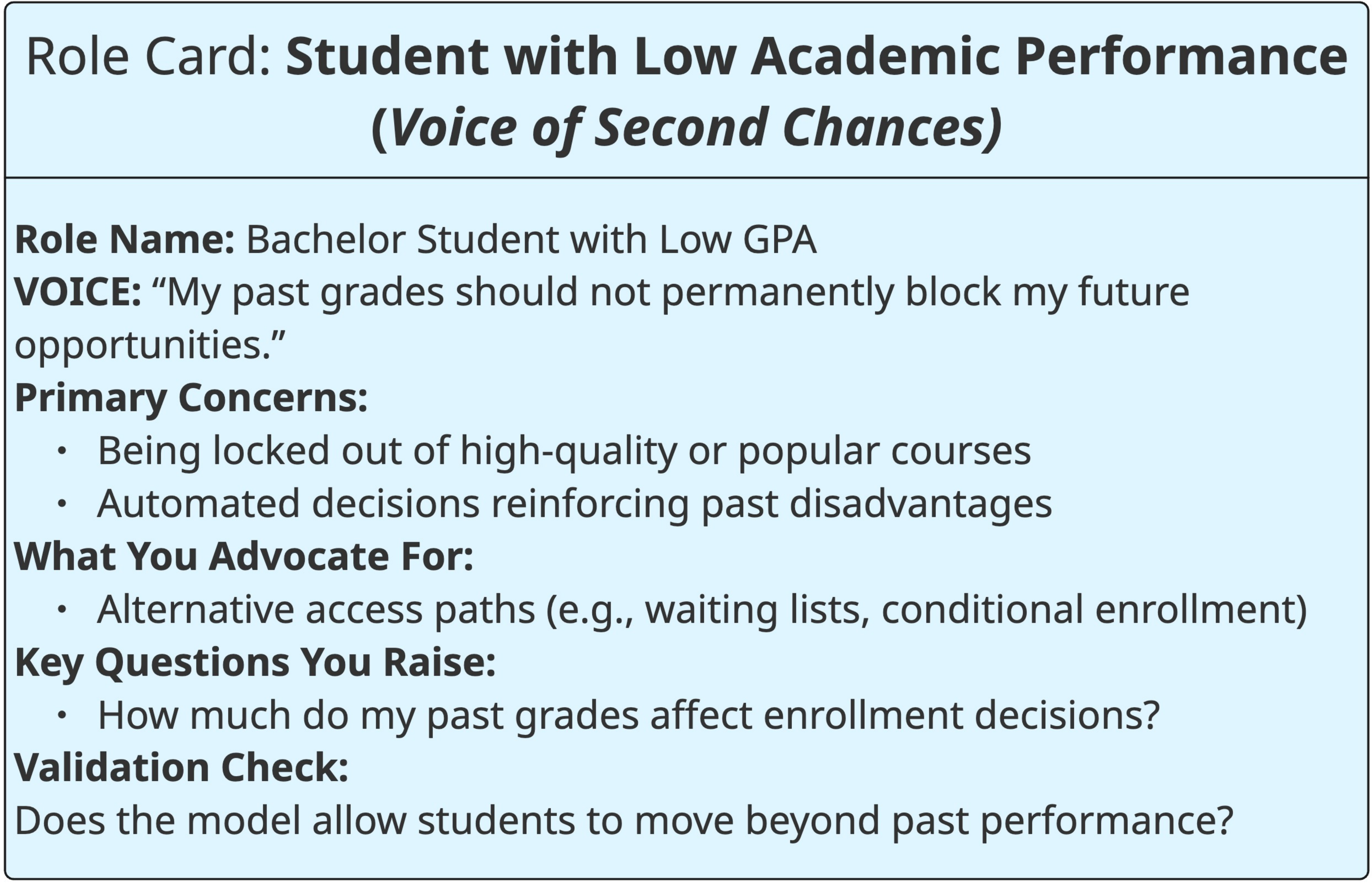}
        \caption{Example Role Card (Voice) from \textbf{the Course Enrolment System} scenario. 
        The card represents the \textit{Voice of Second Chances}, making concerns about 
        grade-based exclusion explicit and traceable during participatory validation.}
        \label{fig:role-card-second-chances}
    \end{subfigure}

    \caption{\GARLIC workshop structure and validation artifacts. 
    Individual Role Cards (right) articulate participant voices within a shared 
    scenario, while the \GARLIC framework (left) structures how these voices are 
    integrated and validated during participatory ER modeling.
    }
    \label{fig:garlic-overview-and-role}
\end{figure*}

The primary contribution of this research is a pedagogical methodology for teaching participatory ER modeling through a structured, hands-on workshop format. The methodology operationalizes the principles of the \ONION participatory modeling framework in an educational setting, making them accessible to participants with diverse backgrounds, including those without prior technical training. Rather than introducing \ONION as an abstract framework, the workshop is designed as a replicable learning experience in which participation, perspective-taking, and validation are central learning outcomes. Figure~\ref{fig:garlic-overview-and-role} provides an overview of the \GARLIC workshop structure and illustrates how participant voices are articulated and carried through the participatory modeling process.

A shared \textbf{Scenario Card} defines the overall design space and encloses two complementary elements: (1) individual \textbf{Role Cards (Voices)}, which articulate participant perspectives, and (2) a \textbf{Participatory Framework}, instantiated through the \ONION stages, which structures collective sense-making and model construction. As illustrated in Figure~\ref{fig:garlic-overview-and-role}\subref{fig:method-overview}, the Scenario Card provides a stable framing context, while Figure~\ref{fig:garlic-overview-and-role}\subref{fig:role-card-second-chances} shows how individual voices are explicitly represented and later used for validation. Together, these elements guide participants from individual viewpoints towards a jointly produced concrete ER model while preserving the traceability of voices throughout the process.

This section details the pedagogical principles underlying the workshop, its core components, and the facilitated design process. 

\subsection{Pedagogical Principles}
The workshop leverages experiential and collaborative learning to teach socio-technical modeling.

\textbf{Learning-by-Doing.} Rather than a lecture, the workshop functions as an active simulation. Participants apply participatory practices to a concrete scenario, following Kolb’s experiential cycle: concrete experience, reflection, conceptualization, and active experimentation \cite{kolb1984experiential}.

\textbf{Perspective-Taking through Role-Playing.} Using \emph{Role Cards (Voices)}, participants advocate for stakeholder perspectives potentially different from their own. This deliberate role-separation surfaces conflicting values and fosters empathy, aligning with design justice and PD principles \cite{costanza2020design}.

\textbf{Iteration and Constructive Failure.} The process explicitly values iteration. Tensions or omissions are treated as critical learning moments rather than errors. By revisiting framework stages, participants learn that iteration is integral to responsible system design.

\textbf{Scaffolding Complexity.} To prevent cognitive overload, ER modeling is decomposed into sequential stages. The Scenario Card provides stability, Role Cards anchor contributions, and the \ONION stages structure collective effort, ensuring meaningful engagement regardless of technical expertise.

\subsection{Workshop Components and Materials}

The workshop relies on a small set of tightly integrated components, each of which corresponds to a distinct element in Figure~\ref{fig:method-overview}:

\begin{itemize}
    \item \textbf{Virtual Whiteboard}  
    The workshop is conducted on a pre-configured digital canvas (e.g., Miro\footnote{\url{https://miro.com/}} or Mural\footnote{\url{https://www.mural.co/}}). The canvas visually mirrors the structure shown in Figure~\ref{fig:method-overview}, with a shared scenario space, areas for individual role-based input, and sections corresponding to the \ONION stages. \ONION conceptualizes participatory modeling as a sequence of five stages: \emph{Observe, Nurture, Integrate, Optimize, and Normalize}. Each corresponding to a distinct type of collaborative activity and level of abstraction.

    \item \textbf{Scenario Card}  
    The Scenario Card defines the shared context, problem space, and overall objective of the workshop. It acts as the outer frame within which all activities occur. The scenario is intentionally concise and relatable, and typically contains an inherent tension (e.g., efficiency vs. safety, access vs. privacy) that motivates discussion and modeling decisions. Throughout the workshop, the Scenario Card serves as a stable reference point against which modeling choices are justified.

    \item \textbf{Role Cards (Voices)}  
    Each participant receives a Role Card representing a specific stakeholder perspective. Each Role Card represents a \emph{voice} in the design process; roles are not personas, 
    but advocacy positions that articulate non-negotiable concerns, priorities, and questions to be carried through the modeling process.

    \item \textbf{Participatory Framework (\ONION)}  
    Participants engage in the workshop from distinct perspectives determined by their assigned roles. To support this process, we provide a structured set of \ONION stage cards tailored to three perspectives: participants, facilitators, and technical experts. Each card type guides its holder through the modeling process while maintaining clear boundaries between stakeholder voices, facilitation responsibilities, and technical decision-making.\footnote{A detailed description of \ONION card types, their roles, and the boundaries between facilitation, technical expertise, and participant voices is available in the \GARLIC repository: \url{https://github.com/mariykadreams/GARLIC/blob/main/ONION\%20cards/ONION\%20cards\%20differences\%20and\%20purposes.md}.}
\end{itemize}

No prior knowledge of ER modeling or database design is required. The only prerequisite is participants’ willingness to engage with the scenario and to represent their assigned roles.

\subsection{The Workshop Process: From Individual Voices to a Shared Model}

The workshop unfolds as a facilitated journey from individual perspective articulation toward the collaborative construction and validation of an ER model. The process emphasizes participation, traceability, and reflection, framing data modeling as a socio-technical activity.

The session begins with the introduction of the \textbf{Scenario Card}, which establishes a common design context and shared objective. For pedagogical purposes, scenarios are simplified to keep attention focused on participatory dynamics rather than domain-specific technical complexity.

Participants then work individually with their assigned \textbf{Role Cards (Voices)}. Each participant documents concerns, expectations, and constraints strictly from their role’s perspective. This stage is intentionally non-evaluative and non-negotiative: its goal is to surface diverse viewpoints without premature convergence or dominance effects.

Through Role Cards, individual perspectives become explicitly traceable. An example of such a Role Card, including an explicit validation check, is shown in Figure~\ref{fig:role-card-second-chances}. This traceability is crucial for later validation. As the workshop progresses through the \ONION stages, participants can examine whether and how their role’s concerns are reflected in the emerging ER model. If a perspective is missing or diluted, the group can identify where it was lost and revisit earlier stages to address the gap.

Throughout the process, the facilitator synthesizes the collaboratively generated materials into a working ER diagram. Participants then engage in role-based validation, using their Role Cards to assess the model’s inclusiveness and representational adequacy. This backward navigation—from model to roles to scenario—encourages reflection on both the artifact and the process that produced it.
The workshop not only results in a concrete ER model but also teaches participants how participatory modeling practices help preserve, reconcile, and validate diverse voices within data system design.

\paragraph*{Facilitator Responsibilities Across Workshop Phases}
\GARLIC makes facilitation explicit and teachable by structuring the session around four card types---Scenario Cards, Role Cards (Voices), ONION Stage Cards---with visual cues (e.g., color) that reduce cognitive load and make transitions between activities legible to novices. The facilitator’s responsibility is therefore twofold: (i) coordinate movement through stages and artifacts, and (ii) protect \emph{voice traceability}, i.e., whether stakeholder positions remain locatable in the evolving model.

\emph{Individual role phase (Role Cards as voice scaffolds).} The facilitator ensures each participant understands that roles are \emph{advocacy positions} rather than personas and that they should argue from the assigned voice, not personal opinion. The main intervention at this stage is supportive: clarify the role’s \texttt{VOICE} (non-negotiable claim), encourage participants to articulate concerns and key questions in their role’s language, and help disengaged participants re-enter by pointing back to the Role Card prompts. The facilitator avoids early convergence; the goal is to surface distinct voices before any shared schema is proposed.

\emph{Group synthesis phase (from voices to shared concepts).} During group discussion and sketching, the facilitator prevents the activity from collapsing into standard requirements gathering by actively maintaining plurality of voices. Concretely, the facilitator (i) invites underrepresented roles to speak, (ii) makes omissions explicit (``Which voice have we not heard from yet?''), and (iii) treats tensions as modeling resources rather than failures. When disagreements arise, the facilitator redirects the group from debating whose view is ``right'' to negotiating what needs to be represented (entities, relationships, attributes, or constraints) so that trade-offs remain traceable.

\emph{ONION stage transitions (Stage Cards as coordination scaffolds).} To support novices, \ONION Stage Cards are used as a script for transitions: each stage makes explicit its goal, the expected participant activity, and the intended output. The facilitator announces the transition criteria (e.g., moving from Observe to Nurture once roles and scenario tension are understood; moving from Nurture to Integrate once perspectives have been articulated and externalized), and explicitly legitimizes backtracking when a voice is lost. This reduces ``black-box'' facilitation by showing students \emph{when} to explore, \emph{when} to articulate, \emph{when} to integrate, and \emph{when} to refine.

\emph{Validation phase (Role Cards reused for participatory validation).} \GARLIC frames validation as a learning activity centered on traceability rather than a search for a single correct diagram. The facilitator prompts each participant to apply the Role Card \emph{validation check} to the current ER model by asking: \emph{``Where is this voice represented in the ER model?''} If a voice cannot be located in an entity, relationship, attribute, or constraint, the participatory process is treated as \emph{incomplete, not incorrect}. The facilitator then guides the group to revisit the relevant stage and adjust the model accordingly. Where applicable, a designated technical specialist supports the Integrate/Normalize steps by translating drafts into a coherent ER diagram and confirming that refinements preserve technical soundness.

\section{Preliminary Studies and Formative Refinements}


To test the teachability and practical usability of \GARLIC, we conducted two formative pilot workshops (5 participants each) using scenarios drawn from our open materials: one centered on a library management system context and one on a community tool shed system. Participants were primarily second- and third-year undergraduate students from non-technical majors. Most had no prior exposure to Entity–Relationship modeling; only one participant had previous experience working on an ER-related task in a team setting. One workshop was conducted online and the other offline. Both were standalone 90-minute sessions designed specifically to test and refine the \GARLIC materials, rather than as part of a formal database course.

Across both pilots, the primary facilitation challenge was preventing ``solutioning'' and domain deep-dives from crowding out the participatory objectives of each card stage. Participants frequently moved prematurely toward defining entities and relationships before fully articulating stakeholder voices during the \emph{Observe} and \emph{Nurture} stages. In several instances, Role Cards were initially treated as descriptive personas rather than advocacy positions, leading to partial neglect of voice constraints during early modeling. Participants also required structured facilitation support to consistently apply the voice-traceability validation check (i.e., explicitly locating each voice in entities, relationships, attributes, or constraints). Without prompting, validation was sometimes interpreted as technical correctness rather than representational inclusion. Nevertheless, all groups were able to progress through the ONION stages and, when prompted, revisit earlier stages to address missing or underrepresented perspectives.

Based on these observations, we introduced several refinements to the workshop framework. First, Role Cards were rewritten with clearer descriptions of the ``VOICE'' as a non-negotiable advocacy position, reducing confusion between personal opinion and assigned perspective. Second, we introduced a leveled scenario progression, beginning with simpler cases involving fewer interacting constraints and gradually moving toward more complex systems with intertwined stakeholder tensions. This scaffolding reduced cognitive overload and enabled participants to internalize the participatory logic before engaging in structurally dense scenarios. 

Post-workshop feedback indicates early signs of conceptual understanding and increased modeling confidence. Participants reported gaining a clearer basic understanding of ER diagrams and increased confidence in constructing models after the workshop. Several participants emphasized that clearly articulated Role Cards helped them think from perspectives different from their own and encouraged hearing ``all voices, not just the loudest ones.'' Participants also reported feeling included in group discussions and valued in the integration process. While these findings are qualitative and exploratory, they suggest that \GARLIC may foster both technical understanding and participatory awareness.

Facilitation proved critical in maintaining participatory integrity. Facilitators intervened primarily in three situations: (1) when discussion drifted into premature structural solutioning; (2) when certain voices became underrepresented; and (3) when validation was reduced to technical correctness rather than voice traceability. Effective prompts included: ``Which voice have we not heard from yet?'', ``Where in the ER model is this concern represented?'', and ``Are we negotiating correctness, or representation?'' Conversely, facilitators deliberately avoided intervening during initial voice articulation to prevent premature convergence. This structured balance between guidance and restraint emerged as a central pedagogical principle refined through the pilot studies.

After the refinments we then ran a short in-class enactment with two small teams (3 participants each): one worked on a student enrollment scenario and one revisited the library system scenario. Because teams were small, each selected three voices, and the instructor alternated as facilitator across groups while giving teams time to deliberate independently. This classroom run further reinforced the importance of explicit transition criteria between stages and lightweight ``traceability checks'' (locating each selected voice in the emerging model) to keep the discussion aligned with the intended learning outcomes (see \autoref{app:library-case} and \autoref{app:enrollment-case}).

\section{Summary and Future Work}
With this work we aimed to address a gap in database education, where ER modeling is typically taught as a purely technical task, despite growing evidence that data models embed social, ethical, and political assumptions. 
By combining scenario framing, role-based perspective taking, collaborative synthesis, and explicit validation of participant voices, the methodology makes visible how individual perspectives shape the resulting ER model. Scenario and Role Cards serve as anchors that preserve participant input throughout the process and allow students to assess whether the participatory goals were achieved.

\GARLIC operationalizes 
core principles of participatory design in an effective classroom methodology for teaching and learning participatory ER modeling. 
While \GARLIC is informed by \ONION, it is not limited to that framework and can be adapted to other participatory approaches. The central aim is to make participatory modeling processes simple, memorable, and easy to validate, without framing participation in terms of correctness or error. Instead, \GARLIC treats missing or misrepresented perspectives as signals for revisiting earlier steps in the process rather than as failures. Through this approach, \GARLIC prepares students to engage with ER modeling as a sociotechnical practice and equips future engineers with practical tools for facilitating inclusive data system design.

We have presented the results of our ongoing research, building on two preliminary studies with learners and two more followup workshops with smaller teams after revising the structure of \GARLIC. Future work includes deeper empirical analysis and refinement of the framework and open-source teaching materials. These activities served as \emph{formative} validation to refine the materials and facilitation scaffolding rather than as a summative evaluation of learning outcomes.

\section{Acknowledgments}

This research was conducted as part of the RAI for Ukraine program (https://r-ai.co/ukraine),
run by the Center for Responsible AI at New York University
in collaboration with Ukrainian Catholic University in Lviv. This
research was supported in part by NSF Awards No. 2520637, 2312930, and 2326193.

\bibliography{sample-ceur}

\clearpage
\appendix

\section{Preliminary Case Study: Library System Tool}
\label{app:library-case}

\begin{figure*}[t]
    \centering
    \includegraphics[width=0.95\linewidth]{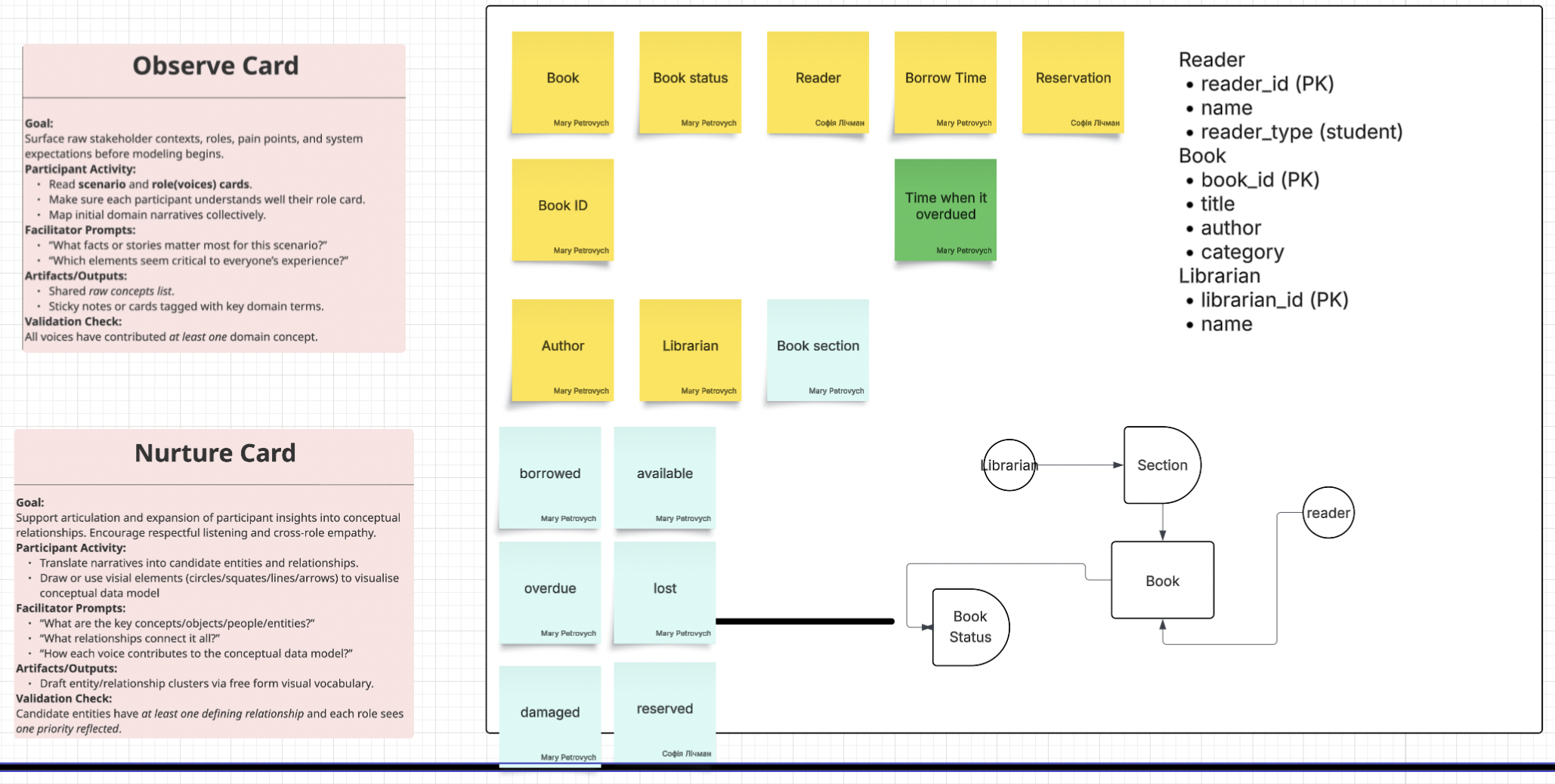}
    \caption{Library case study artifacts from the \textbf{Observe} and \textbf{Nurture} stages. Left: stage card goals/prompts and expected outputs. Center: participant-generated domain concepts and early clusters. Right: an initial sketch linking candidate entities/relationships prior to formalization. Students were allowed to use any visual vocabulary they considered convinient and useful.}
    \label{fig:app-library-observe-nurture}
\end{figure*}

\begin{figure*}
    \centering
    \includegraphics[width=0.95\linewidth]{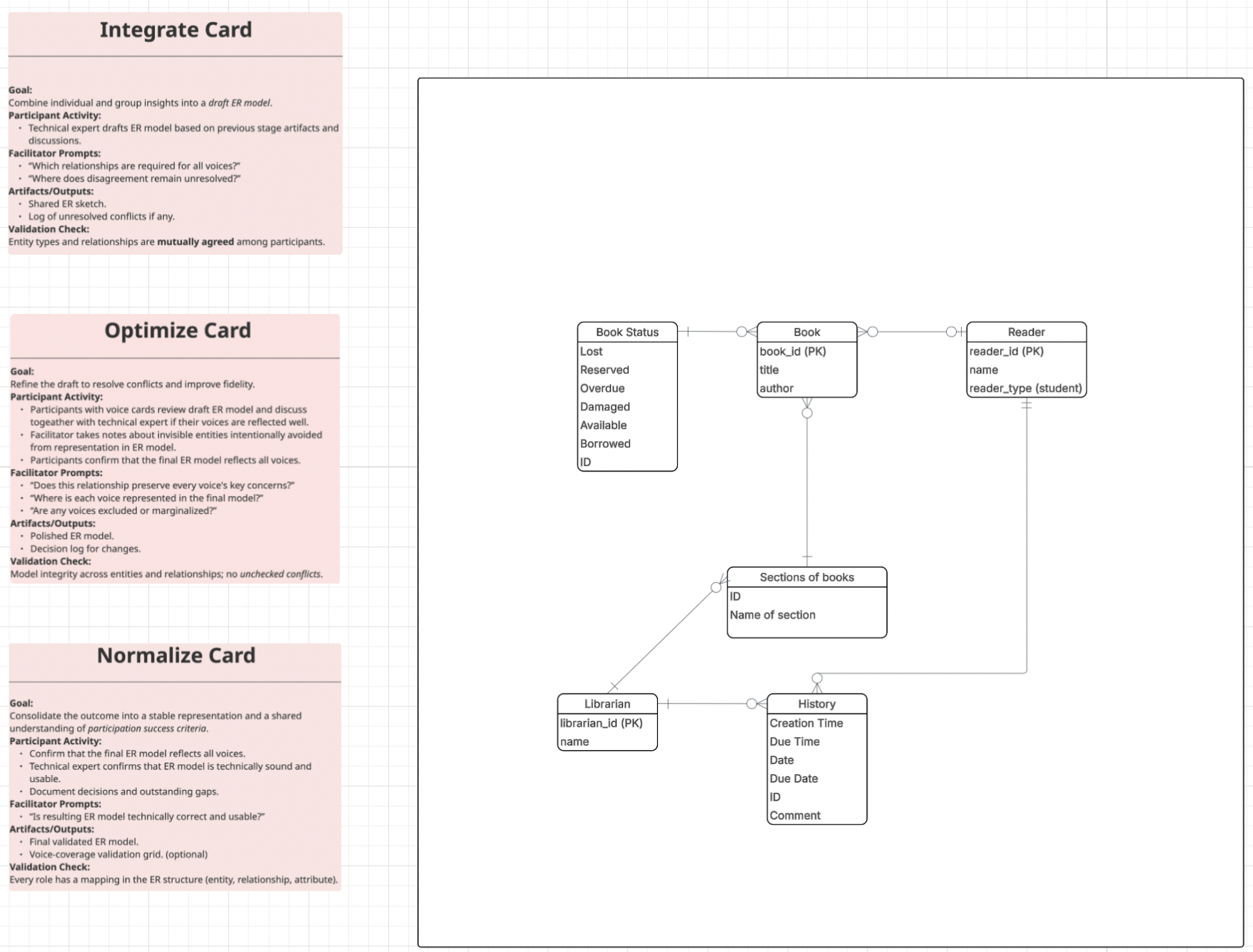}
    \caption{Library case study artifacts from \textbf{Integrate}, \textbf{Optimize}, and \textbf{Normalize}. Left: stage card prompts focusing on unresolved tensions and voice traceability. Right: consolidated ER draft used for role-based validation (mapping each selected voice to entities, relationships, attributes, or constraints).}
    \label{fig:app-library-integrate-normalize}
\end{figure*}

Additionally to first two \GARLIC workshops with 5 participants each (library system and community shed tool), we conducted an additional formative pilot repeating a \textbf{Library System} scenario (3 participants) with a goal to implement corrections which were derived from first two studies. The primary facilitation issue was participants drifting into implementation details (e.g., UI features, policy edge cases) rather than completing the objective of the current stage card. We addressed this by time-boxing each stage and explicitly redirecting discussion to the \emph{card objectives} (outputs) and \emph{voice traceability} checks.

\autoref{fig:app-library-observe-nurture} shows early-stage artifacts from \textbf{Observe} and \textbf{Nurture}: initial domain concepts elicited from the Scenario and Role Cards, along with draft conceptual clusters and tentative relationships. \autoref{fig:app-library-integrate-normalize} shows the subsequent consolidation across \textbf{Integrate}/\textbf{Optimize}/\textbf{Normalize} into a draft ER model used for role-based validation. The draft ER diagram was validated by explicitly answering the validation questions on the selected Role Cards and checking whether each voice could be located in specific entities, relationships, attributes, or constraints. The workshop was time-boxed to 90 minutes.

Participants completed the activity after a theoretical lecture on Entity--Relationship modeling. This enabled them to apply ER concepts during the technical steps and to perform both internal (technical soundness) and external (voice traceability) validation during the session. In small groups, this sometimes required temporarily setting Role Cards aside during technically focused steps (e.g., \textbf{Optimize}) to ensure the draft model remained coherent. In settings with more participants, this burden can be reduced by assigning a dedicated technical-expert role, allowing other participants to maintain continuous voice advocacy throughout the workflow.

\section{Preliminary Case Study: Course Enrollment Scenario}
\label{app:enrollment-case}

\begin{figure*}
    \centering
    \includegraphics[width=0.95\linewidth]{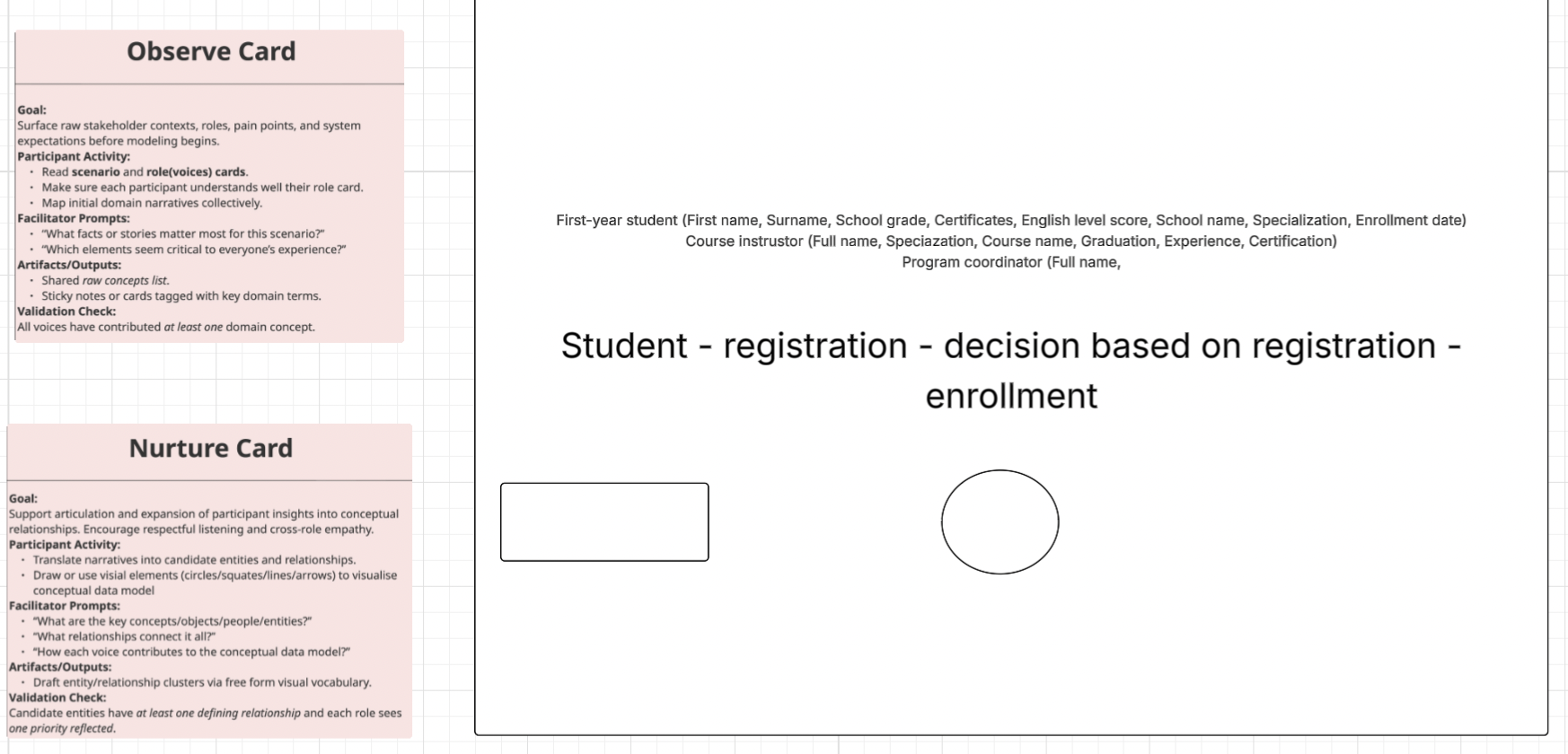}
    \caption{Course Enrollment scenario artifacts from \textbf{Observe} and \textbf{Nurture}. The team recorded a compact set of domain concepts and stakeholder concerns and moved early toward structural representation, illustrating a compressed early-stage workflow in a small-group classroom setting.}
    \label{fig:app-enrollment-observe-nurture}
\end{figure*}

\begin{figure*}
    \centering
    \includegraphics[width=0.95\linewidth]{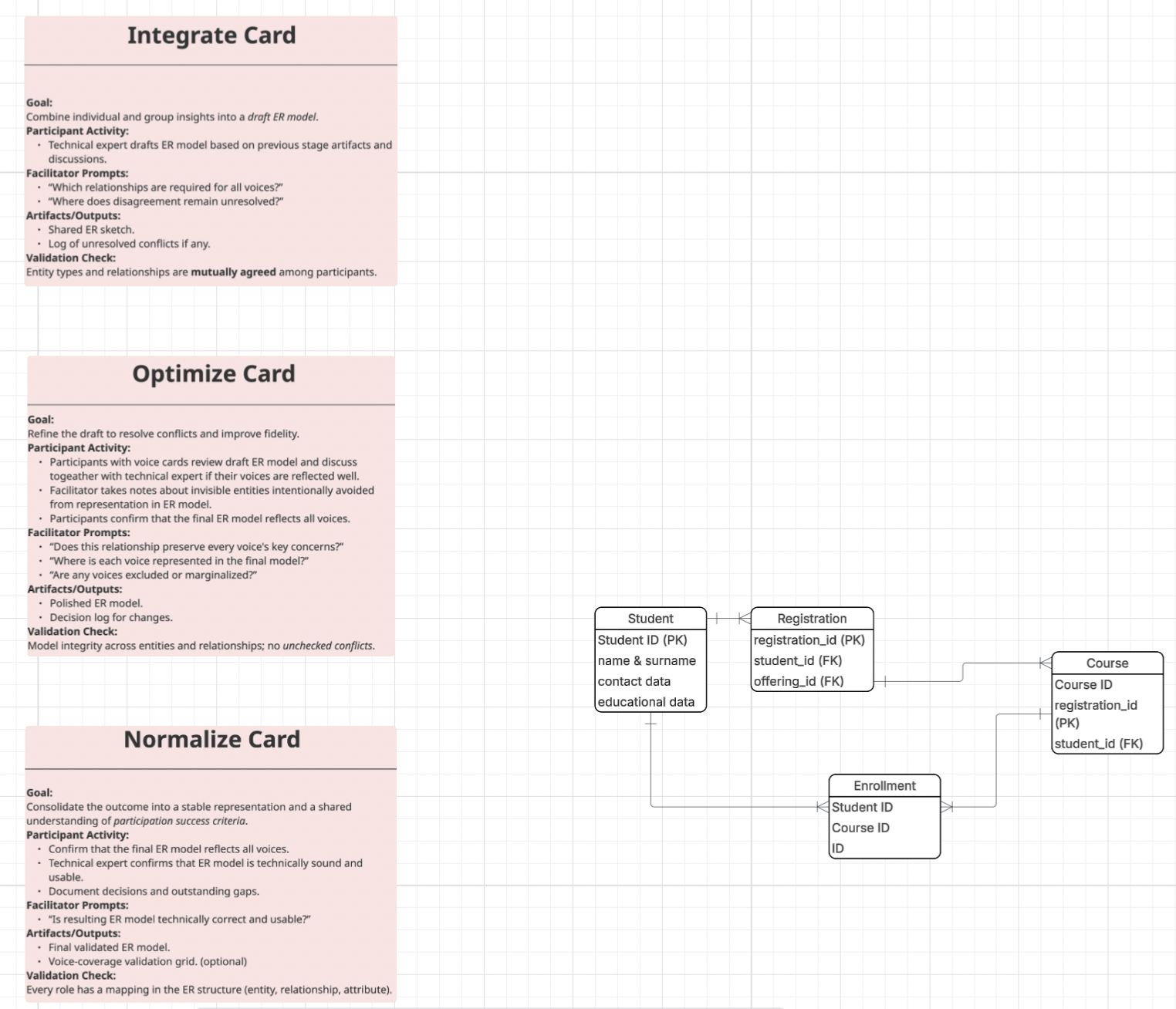}
    \caption{Course Enrollment scenario artifacts from \textbf{Integrate}/\textbf{Optimize}/\textbf{Normalize}. The team produced a draft ER model and refined it through technical validation (external validation). After voice validation (internal validation) it was decided that the ER model needs refinment and returning to previous stages needed.}
    \label{fig:app-enrollment-integrate-normalize}
\end{figure*}

We ran an additional short in-class enactment of \GARLIC using a \textbf{Course Enrollment System} scenario with a small team (3 participants), who therefore selected three voices. Relative to the library case, this group generated fewer intermediate artifacts and adopted a more ``direct-to-structure'' visualization style: rather than elaborating extensive concept clusters during the early stages, they transitioned quickly to defining candidate entities and relationships and then refined structure during later stages. This variation is instructive for classroom deployment, as small groups and tight time constraints often compress \textbf{Observe}/\textbf{Nurture} (\autoref{fig:app-enrollment-observe-nurture}, \autoref{fig:app-enrollment-integrate-normalize} ) and concentrate effort in \textbf{Integrate}/\textbf{Optimize}/\textbf{Normalize}. To maintain alignment with \GARLIC objectives, the facilitator time-boxed discussion, redirected implementation-level digressions back to stage-card prompts, and reintroduced Role Card validation checks before finalizing the ER draft. Because time was limited, the team did not finalize an ER diagram that met the voice-traceability validation criterion; this was turned into a follow-up exercise in which students returned to earlier stages to elicit missing concerns and incorporate the corresponding entities, relationships, attributes, or constraints into the model.

\end{document}